\documentclass[runningheads]{llncs}
\usepackage[T1]{fontenc}
\usepackage{newtxtext}       %
\usepackage[varvw]{newtxmath}       

\usepackage{graphicx}
\usepackage{amsmath}
\usepackage[acronym]{glossaries-extra}
\usepackage{bm}
\usepackage{bbding}
\usepackage{adjustbox}
\usepackage{tikz}
\usetikzlibrary{trees,positioning}
\usepackage{subcaption}
\usepackage{tabularray}
\usepackage{xurl}
\usepackage[hidelinks]{hyperref}
\usepackage{float}
\usepackage[linesnumbered,ruled,vlined]{algorithm2e}
\usepackage{algpseudocode}
\usepackage{listings}
\usepackage{xcolor}
\usepackage{ifsym}
\usepackage[shortlabels]{enumitem}
\usepackage{booktabs}

\setabbreviationstyle{long-short}
\newabbreviation{he}{HE}{Homomorphic Encryption}
\newabbreviation{fhe}{FHE}{Fully Homomorphic Encryption}
\newabbreviation{bgv}{BGV}{Brakerski-Gentry-Vaikuntanathan}
\newabbreviation{bfv}{BFV}{Brakerski-Fan-Vercauteren}
\newabbreviation{ckks}{CKKS}{Cheon-Kim-Kim-Song}
\newabbreviation{tfhe}{TFHE}{Fast Fully Homomorphic Encryption over the Torus}
\newabbreviation{nist}{NIST}{National Institute of Standards and Technology}
\newabbreviation{pqc}{PQC}{Post-Quantum Cryptography}
\newabbreviation{ehr}{EHR}{Electronic Health Record}
\newabbreviation{ntt}{NTT}{Number Theoretic Transform}
\newabbreviation{fft}{FFT}{Fast Fourier Transform}
\newabbreviation{fafft}{FAFFT}{Frobenius Additive FFT}
\newabbreviation{rlwe}{R-LWE}{Ring LWE}
\newabbreviation{lwe}{LWE}{Learning With Errors}
\newabbreviation{mlwe}{M-LWE}{Module LWE}
\newabbreviation{fhew}{FHEW}{Fastest Homomorphic Encryption in the West}
\newabbreviation{rns}{RNS}{Residue Number System}
\newabbreviation{cbd}{CBD}{Centered Binomial Distribution}
\newabbreviation{hess}{HESS}{Homomorphic Encryption Security Standard}
\newabbreviation{crt}{CRT}{Chinese Remainder Theorem}
\newabbreviation{ct}{CT}{Cooley-Tukey}
\newabbreviation{gs}{GS}{Gentleman-Sande}
\newabbreviation{ecg}{ECG}{Electrocardiogram}
\newabbreviation{kem}{KEM}{Key-Encapsulation Mechanism}
\newabbreviation{mlkem}{ML-KEM}{Module-Lattice-Based Key-Encapsulation Mechanism}
\newabbreviation{mldsa}{ML-DSA}{Module-Lattice-Based Digital Signature Algorithm}
\newabbreviation{dft}{DFT}{Discrete Fourier Transform}
\newabbreviation{intt}{INTT}{Inverse Number Theoretic Transform}
\newabbreviation{sdnn}{SDNN}{Standard Deviation of NN Intervals}
\newabbreviation{nn}{NN}{Normal-to-Normal}
\newabbreviation{pwm}{PWM}{Point-Wise Multiplication}
\newabbreviation{prf}{$\mathsf{PRF}$}{Pseudorandom Function}
\newabbreviation{xof}{$\mathsf{XOF}$}{eXtendable-Output Function}
\newabbreviation{iot}{IoT}{Internet of Things}

\begin{document}
\title{An Agile Homomorphic Encryption and ML-KEM Scheme for Privacy-Preserving Health Data Processing}
\subtitle{(Extended Version -- Lightweight Cryptography for Security and Privacy 2026)\footnote{Please cite the workshop version of this paper \cite{REK}, published at LigthSec 2026.}}
\titlerunning{An Agile HE and ML-KEM Scheme for Health Data Processing}
%
\author{Daniel Resende\inst{1} \Envelope \orcidID{0009-0005-6382-8731} \and 
Nadia El-Mrabet\inst{1}\orcidID{0000-0003-3840-584X} \and
Ulrich Kühne\inst{2}\orcidID{0000-0002-0855-8223}}

\authorrunning{D. Resende et al.}
%
\institute{Mines Saint-Etienne, CEA-Leti, Centre CMP, Equipe commune SAS, F-13541, Gardanne, France\\
\email{\{daniel.resende, nadia.el-mrabet\}@emse.fr} \and
Télécom Paris, Institut polytechnique de Paris, Palaiseau, France \email{ulrich.kuhne@telecom-paris.fr}}
\maketitle              
\begin{abstract}
The storage and processing of health data raise serious privacy concerns; therefore, the data must be encrypted. \gls{fhe} allows computations on encrypted data, which makes it a very promising technology in this context. However, although the efficiency of \gls{fhe} algorithms has improved in recent years, they remain very complex. The security of several prominent \gls{fhe} schemes is based on variants of the \gls{lwe} problem, which is also the basis of \gls{mlkem}, the new post-quantum standard for asymmetric encryption. In this paper, we propose an agile cryptographic scheme -- combining \gls{mlkem} and BGV -- that allows efficient implementations in an embedded use case. We describe the parameter selection for the proposed scheme and show its feasibility with a proof-of-concept implementation.

\end{abstract}
\section{Introduction}
\label{sec:introduction}

The use of digital health data offers new opportunities for medical research and improved care, but it also poses an increasing risk to the privacy of the concerned patients. News about data breaches of sensitive health data has become common. The number of reported incidents affecting at least 500 patients in the US has grown from around 200 in 2010 to more than 700 in 2024~\cite{alderBiggestHealthcareData2025}. Even worse, almost $70\,\%$ of the US population was affected by these breaches in 2024. In order to manage vast amounts of data, hospitals and other health providers are using cloud storage as a scalable solution. While this offers clear advantages regarding availability and efficiency, it also exposes the stored data to attacks~\cite{diaz1YearLater2025}. 

Another source of health data are portable embedded devices, ranging from consumer electronics such as fitness trackers or smart watches to specialized medical devices such as heart rate monitors or implanted glucose trackers, some of which are connected to the internet in order to report or share the collected data. The possible consequences of carelessly sharing such data have been reported e.g.~in~\cite{slyUSSoldiersAre2018}, when the position of secret military bases was revealed by GPS data from soldiers’ fitness trackers. 
At the same time, more and more countries -- for example in Europe~\cite{europeancommission2025ehds} -- are transitioning to the official use of \glspl{ehr}. 
As the study in~\cite{vonkalckreuthTrustworthinessElectronicHealth2025} reveals, the large scale adoption of such systems by patients critically depends on their \emph{trustworthiness}. This shows that security is crucial in any application that handles health data, and that storing health records in plain text should be avoided in the first place.
The proper use of cryptography is paramount.

Although there are many robust cryptographic algorithms, their use does not come without friction.
We are currently in a transitional phase, where some widely adopted cryptographic standards are being deprecated due to the potential threat of quantum computers.
While new (post-quantum) algorithms have already been standardized -- such as the \gls{mlkem}~\cite{nist2024_fips203} -- they are more complex than the ones they replace, and robust and efficient implementations are currently under development.

Furthermore, in a typical healthcare scenario, there can be multiple parties that should have legitimate access to the recorded data, including doctors and medical staff, pharmacies, research institutions, and the patient herself. This makes the use of symmetric cryptography difficult, since the secret key needs to be distributed in a secure way. 
Even when the data is stored encrypted, it needs to be temporarily decrypted for any subsequent processing or aggregation, exposing it to potential attackers.

In this context, \gls{he} is a very attractive solution. It allows computations to be performed on encrypted data \emph{without decrypting} it. This offers the possibility to do computations on a cloud server, while maintaining full control and ownership over the data, and without needing to trust the server. The great potential of \gls{he} in health applications has been explored e.g.~in~\cite{munjalSystematicReviewHomomorphic2023,SecureHealthcareData2025}. A drawback of current \gls{he} schemes is the computational complexity of the cryptographic operations. While the homomorphic computations can be carried out on a powerful cloud server, at least encryption and decryption must take place on the user's end device. In the scheme proposed in~\cite{SecureHealthcareData2025}, the key generation is taken care of by a dedicated server. This limits the user's trust in the scheme. It has yet to be shown that \gls{he} can be used in a realistic embedded setting, where a low power device has full control over both key generation and encryption. 

In this paper, we propose an agile cryptographic scheme for exactly this application scenario. In particular, we leverage the fact that several interesting \gls{fhe} algorithms are based on the same -- or a very similar -- mathematical framework as \gls{mlkem}, namely \gls{lwe}. With this observation in mind, we select algorithmic parameters in such a way as to maximize the synergies in the implementation of both \gls{mlkem} and \gls{fhe}, in order to reduce the code size of the cryptographic implementation. 
The contributions of this paper are the following:
\begin{itemize}
    \item An embedded use case scenario for an agile cryptographic scheme.
    \item A systematic review of existing \gls{fhe} algorithms and their suitability for the use case.
    \item The definition and parameterization of a cryptographic scheme combining \gls{mlkem} and \gls{fhe}. 
    \item A proof of concept implementation for the above scheme.
\end{itemize}

The remainder of this paper is organized as follows:
We will start with a presentation of the use case scenario in \autoref{sec:usecase}. The necessary background for the used cryptographic algorithms is given in \autoref{sec:background}. Based on these definitions, we select a suitable algorithm and its parameters in \autoref{sec:schema-selection}. Possible optimizations are discussed in \autoref{sec:bgv-optimization}. Finally, a proof-of-concept implementation is presented in \autoref{sec:implementation} before concluding the paper in \autoref{sec:conclusion}.

\vspace{-0.15cm}
\section{Use Case: Average Heart Rate Computation}
\label{sec:usecase}
\vspace{-0.25cm}
The objective is to protect medical research and data analysis in healthcare monitoring. 
More specifically, this involves ensuring the confidentiality and security of data transmission and storage, particularly regarding personal health records stored on wearable devices, also called \glspl{ehr}.

\vspace{-0.25cm}
\subsection{System Overview}
\label{subsec:system}

We chose this use case due to the abundance of available studies and the ease of access to the required sensors, which are integrated into widely available connected devices (such as smartwatches and smartphones) or medical devices (like blood pressure monitors). 
A concrete example would be an athlete who wants to track their health on a mobile app and a smartwatch, with direct access to this data for a healthcare professional.

\begin{figure}
    \vspace{-0.35cm}
    \centering
    \includegraphics[width=0.9\textwidth]{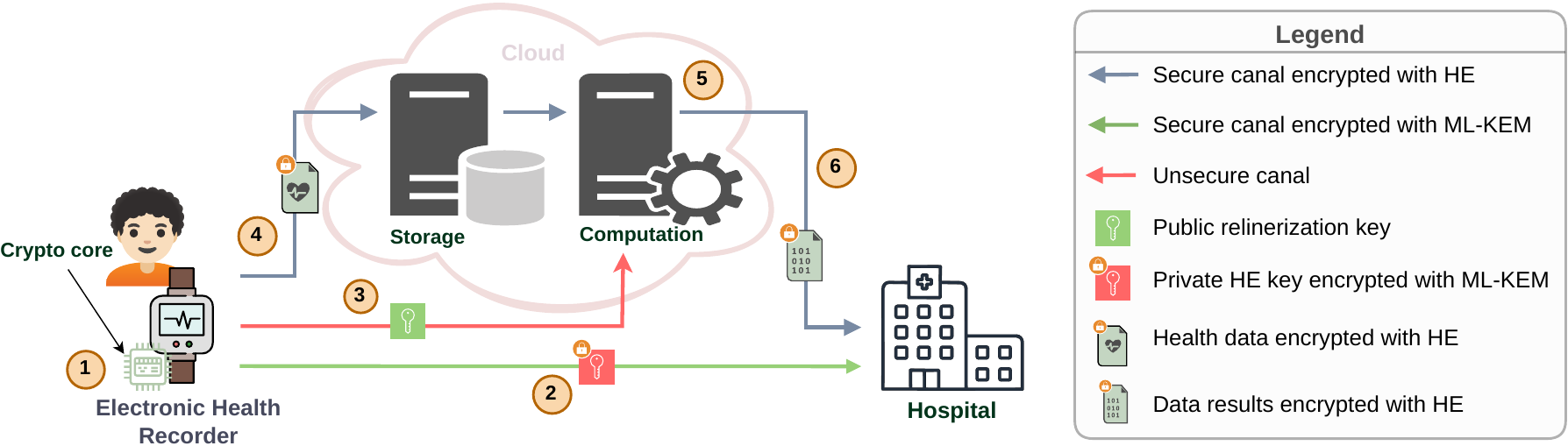}
    \caption{Use case schema.}
    \label{fig:health-smartwatch}
    \vspace{-.75cm}
\end{figure}

As depicted in~\autoref{fig:health-smartwatch}, the concept is to leverage health data (discussed in \autoref{subsec:data}) from an \gls{ehr}, specifically a smartwatch, for processing or statistical analysis on a distant server or cloud. This ensures the protection and security of the data throughout the server-side processing.
Once the data has been processed, the results are securely transmitted, either to the patient or directly to the healthcare provider, such as a doctor or hospital.

Given the constraints of the \gls{iot} environment (limited memory, reduced surface area), we favor a centralized approach: data is stored and encrypted on the server, accessible for statistical analysis at any time, without requiring processing on the devices.

To achieve this, we propose utilizing homomorphic encryption, as depicted in \autoref{fig:health-smartwatch}, to encrypt sensitive medical information.

Furthermore, to enable communication between different equipment (such as the \gls{ehr}, smartphones, and servers), it is essential to establish and exchange cryptographic keys, including homomorphic keys, to ensure the integrity and protection of these critical assets.

The goal is to consolidate cryptographic functions into a single \textit{crypto-core} that combines key exchange, homomorphic encryption and more traditional symmetric encryption\footnote{We will not address this point of symmetric encryption in this article.} (such as Ascon~\cite{NIST:SP800-232}) for message exchange (for data that do not require confidentiality at the server level, because, as we will see in~\autoref{subsec:he}, it is more costly).

This data contains information about the electrical activity of the patient's heart, recorded over 24 hours. In order to demonstrate the feasibility of our concept, we selected average heart rate calculation as our case study. To compute the average heart rate we use information extracted from the annotation file of the ECG.
The annotation file provides information related to cardiac contraction for each beat and the temporal distance (toc) between consecutive beats. The toc values between consecutive beats will be used to calculate the average heart rate.

\paragraph{Process flow described in \autoref{fig:health-smartwatch}:}
\label{para:process-flow}
\begin{enumerate}
\item Key Generation: The crypto-core will generate the homomorphic key pair.
\item Key Exchange (Optional): We exchange the private homomorphic key, using ML-KEM encryption, to allow a third party (such as a doctor or hospital) to decrypt the results returned by the server\footnote{The keys of the KEM can be managed in different ways; in particular, one should avoid a centralized PKI operated by the same provider as the computing server.}.
\item Key Exchange: Moreover, the relinearization key is sent to the server, as it is necessary after a multiplication.
\item Encryption: Data generated by the device (e.g., a smartwatch) is encrypted and sent to the server.
The data is stored encrypted in the database.
\item Computation: When a user or third party makes a statistical request, the server performs arithmetic operations (such as addition or multiplication) on the encrypted data, resulting in an encrypted output.
\item Decryption: The server sends the encrypted results back to the user or third party, who can decrypt them using their private homomorphic key.
\end{enumerate}

\subsection{Analyzed Data}
\label{subsec:data}

Kocabas et al.~\cite{6657078Kocabas} describe the computation of the average heart rate using a sample \gls{ecg} annotation file, such as the database provided by~\cite{5628067Couderc}.
In our case, we only focus on descriptive statistical functions such as summation, mean, variance, or standard deviation.
Given a signal sampling frequency of $f_s = 1{,}000$~Hz, each value $toc_i$ corresponds to the time interval between two successive heartbeats, measured in units of samples.
For $N$ beats, the values are denoted as $toc_1, \ toc_2, \ \dots,\ toc_N$.

\begin{enumerate}
    \item \textbf{Sum of Intervals:} $S = \sum_{i=1}^{N} toc_i$.\\
    \item \textbf{Mean RR Interval (in seconds):} The RR intervals represent the time between each detected heartbeat, measured from peak (R) to peak (R).$$\overline{RR} = \frac{S}{N} \times \frac{1}{f_s} = \frac{S}{N} \times 0.001.$$
    
    \item \textbf{Heart Rate (in beats per minute, bpm):} $$\texttt{HR}_{\text{bpm}} = \frac{60}{\overline{RR}} = 60 \times \frac{f_s \cdot N}{S} = 60{,}000 \times \frac{N}{S}.$$
    Since $N$ is a public value, we only need homomorphic operations to compute $S$; after decryption, the heart rate is obtained as $60{,}000 \times N / S$.\\

    \item \textbf{\gls{sdnn}:} The same idea applies to \gls{sdnn}, the overall standard deviation of the \gls{nn} intervals between consecutive normal beats over a given period~\cite{Shaffer}.
    It is defined as: $$\texttt{SDNN} = \sqrt{\frac {1}{N-1}\sum_{i=1}^N ((toc_i - toc_{i + 1}) - \overline{RR})^2}.$$
\end{enumerate}

We just need to perform homomorphic operations on the sum. For each encrypted $toc_i, toc_{i + 1}$, there is a multiplicative depth of one (due to squaring).
\section{Background}
\label{sec:background}

In this section, we define the cryptographic notions of \gls{kem} and \gls{he}, which are used in our use case. We have opted for the latest \gls{pqc} \gls{nist} standard for key exchange, \gls{mlkem}. Recent \gls{he} schemes and \gls{mlkem} are based on variants of the \gls{lwe} problem.

\begin{definition}[\gls{lwe} Distribution]
Let $n \geq 1$, $q$ a prime, $\chi_e$ an error distribution on $\mathbb{Z}_q^n$, and $s \in \mathbb{Z}_q^n$.
The \textbf{\gls{lwe} distribution} $A_{s,\chi_e}$ on $\mathbb{Z}_q^n \times \mathbb{Z}_q$ is defined as:
$$A_{s,\chi_e} = \left\{(a,\, a \cdot s + e \bmod q) \mid a \xleftarrow{\$} \mathbb{Z}_q^n,\ e \leftarrow \chi_e \right\}.$$
This gives rise to two hard problems:
\begin{itemize}
    \item \textbf{Search \gls{lwe}:} Given $m$ samples from $A_{s,\chi_e}$, recover $s$.
    \item \textbf{Decision \gls{lwe}:} Given $m$ samples from either $A_{s,\chi_e}$ (with $s \xleftarrow{\$} \mathbb{Z}_q^n$) or $\mathcal{U}(\mathbb{Z}_q^n \times \mathbb{Z}_q)$, no PPT adversary $\mathcal{A}$ can distinguish them with non-negligible advantage:
    $$\left|\Pr\!\left[\mathcal{A}(A_{s,\chi_e}^m) = 1\right] - \Pr\!\left[\mathcal{A}(\mathcal{U}^m) = 1\right]\right| \leq \varepsilon(n).$$
\end{itemize}
Both variants are conjectured to be equivalent~\cite{regev2024latticeslearningerrorsrandom}.
\end{definition}

\vspace{-0.15cm}

\begin{definition}[\gls{rlwe} Distribution]
Let $n \geq 1$ be a power of two, $q$ a prime, $R_q = \mathbb{Z}_q[X]/(X^n+1)$, $\chi_e$ an error distribution on $R_q$, and $s \in R_q$.
The \textbf{\gls{rlwe} distribution} $A_{s,\chi_e}$ on $R_q \times R_q$ is defined as:
$$A_{s,\chi_e} = \left\{(a,\, a \cdot s + e \bmod q) \mid a \xleftarrow{\$} R_q,\ e \leftarrow \chi_e \right\}.$$\vspace{-.25cm}

Given $m$ samples from either $A_{s,\chi_e}$ (with $s \xleftarrow{\$} R_q$) or $\mathcal{U}(R_q \times R_q)$, no PPT adversary $\mathcal{A}$ can distinguish them with non-negligible advantage:
$$\left|\Pr\!\left[\mathcal{A}(A_{s,\chi_e}^m) = 1\right] - \Pr\!\left[\mathcal{A}(\mathcal{U}^m) = 1\right]\right| \leq \varepsilon(n).$$
\end{definition}

\vspace{-0.6cm}
\subsection{\gls{mlkem}}
\label{subsec:ml-kem}

\gls{mlkem} is a quantum-resistant \gls{kem} whose security reduces to the \gls{mlwe} problem, a generalization of \gls{rlwe}, itself a variant of \gls{lwe}, which was introduced by Regev in 2005~\cite{10.1145/1060590.1060603} and is reducible to standard lattice problems.
It consists of three steps: key generation, encryption, and decryption, parametrized by a prime modulus $q$, a dimension $n$ over $\mathcal{R}_q^k = \bigl(\mathbb{Z}_q[X]/(X^n + 1)\bigr)^k$, noise parameters $\eta_1, \eta_2$, and $k \in \{2, 3, 4\}$ for the desired security level (cf.~\autoref{tab:parameter_ml_kem}).
We refer the reader to~\cite{nist2024_fips203} for further details.

\vspace{-0.5cm}
\begin{table}[ht]
\centering
    \caption{Parameter sets for \gls{mlkem}}
    \label{tab:parameter_ml_kem}
    \setlength{\tabcolsep}{6pt}
    \resizebox{\textwidth}{!}{\begin{tabular}{lccccccc}
    \toprule
    Parameter set & \gls{nist} Security Level & \(n\) & \(q\) & \(k\) & \(\eta_1\) & \(\eta_2\) & \((d_u, d_v)\) \\
    \midrule
    \text{ML-KEM-512}  & 1 & 256 & 3329 & 2 & 3 & 2 & (10, 4) \\
    \text{ML-KEM-768}  & 3 & 256 & 3329 & 3 & 2 & 2 & (10, 4) \\
    \text{ML-KEM-1024} & 5 & 256 & 3329 & 4 & 2 & 2 & (11, 5) \\
    \bottomrule
    \end{tabular}}
\end{table}

\vspace{-0.25cm}

\paragraph{\textbf{Key Generation.}}
\label{para:keygen}
A private key $s\in \mathcal{R}_q^k$ and a secret error $e \in \mathcal{R}_q^k$ are sampled from binomial distributions.
The public key $(A, t)$ is calculated by the operation $t := A \cdot s + e \in \mathcal{R}_q^k$ where the matrix $A \in \mathcal{R}_q^{k \times k}$ is sampled uniformly at random.
\vspace{-0.25cm}
\paragraph{\textbf{Encryption.}}
\label{para:ml-kem-encryption}
Given a message $m \in \{0,1\}^{256}$, we sample from binomial distributions a random vector $r\in \mathcal{R}_q^k$ and two errors $e_1\in \mathcal{R}_q^k$ and $e_2\in \mathcal{R}_q$.
The ciphertext $(u, v)$ is computed as follows:
\begin{equation}
u := A^T \cdot r + e_1 \in \mathcal{R}_q^k, \quad v := t^T \cdot r + e_2 + \mathsf{Decompress}_q(m, 1) \in \mathcal{R}_q
\end{equation}
  
\paragraph{\textbf{Decryption.}}
\label{para:ml-kem-deryption}
Let $(u, v)$ be a ciphertext and $s$ the private key.
To decrypt the message, we compute the operation $m := \mathsf{Compress}_q\left(v - s^T \cdot u, 1\right).$
\vspace{-0.25cm}
\paragraph{\textbf{Compression/Decompression.}}
\label{para:ml-kem-compress}

The functions $\mathsf{Compress}_q(x,\, d) = \left\lfloor \frac{2^d}{q} \cdot x \right\rceil \bmod 2^d$ and $\mathsf{Decompress}_q(x,\, d) = \left\lfloor \frac{q}{2^d} \cdot x \right\rceil$ perform a compression (resp. decompression), reducing the representation of an element of $\mathbb{Z}_q$ to $d < \lceil \log_2 q \rceil$ bits in order to minimize the size of the transmitted ciphertexts~\cite{nist2024_fips203}.

\subsection{\glsxtrlong{he}}
\label{subsec:he}

In 1978, Rivest, Adleman and Dertouzos~\cite{Rivest1978} described the first \gls{he} scheme, based on the RSA problem, which supports homomorphic multiplication.
The goal of \gls{he} is to evaluate (or compute) on encrypted data, for instance on a cloud server, without decrypting it. 
This ensures data confidentiality compared to traditional encryption, which requires decryption before computation.
Subsequently, some proposals for partial \gls{he} (i.e. for homomorphic addition or multiplication) have been made using the problems of ElGamal~\cite{1057074}, Diffie--Hellman~\cite{1055638} and Paillier~\cite{10.1007/3-540-48910-X_16}.
Afterwards, in 2009, Gentry~\cite{10.1145/1536414.1536440} introduced the first \gls{fhe} scheme, which allows both homomorphic operations to be computed and is based on lattice problems.

\begin{definition}[\glsxtrlong{he}]
A cryptosystem $(\mathsf{Enc}, \mathsf{Dec})$ over a plaintext space $\mathcal{M}$
and a ciphertext space $\mathcal{C}$ is \emph{homomorphic} if there exist binary operations
$\circ$ on $\mathcal{M}$ and $\star$ on $\mathcal{C}$ such that \vspace{-0.2cm}$$\mathsf{Dec}\!\left(\mathsf{Enc}(m_1) \star \mathsf{Enc}(m_2)\right)
    = m_1 \circ m_2,
    \qquad \forall\, m_1, m_2 \in \mathcal{M}.$$
In particular:
\begin{itemize}
    \item \textbf{Additive homomorphism:}
          $\mathsf{Dec}(\mathsf{Enc}(m_1) + \mathsf{Enc}(m_2)) = m_1 + m_2$.
    \item \textbf{Multiplicative homomorphism:}
          $\mathsf{Dec}(\mathsf{Enc}(m_1) \cdot \mathsf{Enc}(m_2)) = m_1 \cdot m_2$.
\end{itemize}
\end{definition}
\vspace{-0.25cm}

\begin{definition}[\glsxtrlong{fhe}]
A \gls{fhe} scheme is a quadruple of algorithms
$(\mathsf{KeyGen}, \mathsf{Enc}, \mathsf{Dec}, \mathsf{Eval})$
defined over a plaintext space $\mathcal{M}$, a ciphertext space $\mathcal{C}$,
and a class of admissible functions $\mathcal{F}$.
It consists of the following algorithms:
\vspace{-0.15cm}
\begin{itemize}
  \item $\mathsf{KeyGen}(\lambda) \to (\mathsf{pk}, \mathsf{sk})$: takes as input a security parameter $\lambda \in \mathbb{N}$ (written in unary) and outputs a public key $\mathsf{pk}$ and a secret key $\mathsf{sk}$.
  \item $\mathsf{Enc}_{\mathsf{pk}}(m) \to c$: takes 
  a plaintext $m \in \mathcal{M}$ and outputs a ciphertext $c \in \mathcal{C}$.
  \item $\mathsf{Dec}_{\mathsf{sk}}(c) \to m$: takes 
  a ciphertext $c \in \mathcal{C}$ and outputs a plaintext $m \in \mathcal{M}$.
  \item $\mathsf{Eval}_{\mathsf{pk}}(f, c_1, \dots, c_n) \to c_f$: takes 
  a function $f \in \mathcal{F}$ and ciphertexts $c_1, \dots, c_n \in \mathcal{C}$, and outputs a ciphertext $c_f \in \mathcal{C}$.
\end{itemize}
\vspace{-0.25cm}
The scheme is said to be \emph{correct} if, for all $f \in \mathcal{F}$, all $n \in \mathbb{N}$, and all $m_1, \dots, m_n \in \mathcal{M}$,
\[\mathsf{Dec}_{\mathsf{sk}}\!\left(\mathsf{Eval}_{\mathsf{pk}}\!\left(f, \mathsf{Enc}_{\mathsf{pk}}(m_1), \dots, \mathsf{Enc}_{\mathsf{pk}}(m_n)\right)\right) = f(m_1, \dots, m_n).\]
\vspace{-0.75cm}
\end{definition}

\section{Scheme Selection}
\label{sec:schema-selection}
\vspace{-0.25cm}
In this section, we detail the selection of the \gls{he} scheme that most closely resembles \gls{mlkem}.
The candidate schemes \gls{bfv}, \gls{bgv}, \gls{fhew}, \gls{tfhe}, and \gls{ckks} are all included in popular homomorphic libraries such as SEAL~\cite{sealcrypto}, OpenFHE~\cite{OpenFHE}, HElib~\cite{Halevi2014AlgorithmsIH}, or HEAAN~\cite{cheon2017homomorphic}.
To guide our selection, we compare their arithmetic structures in \autoref{tab:arith-comparison}, focusing on three criteria: \textbf{(1)} plaintext space compatibility with \gls{mlkem}, \textbf{(2)} ciphertext space structure, and \textbf{(3)} hardware implementation cost.

\begin{table}[H]
    \vspace{-.5cm}
    \centering
    \caption{Comparison of arithmetic aspects of cryptographic schemes.}
    \label{tab:arith-comparison}
    \resizebox{\textwidth}{!}{\begin{tblr}{
        width=0.9\textwidth,
        colspec = {lccl},
        row{1} = {font=\bfseries},
        rowsep=1pt
    }
    \hline
    Schemes & Plaintext & Ciphertext & Security Assumption \\
    \hline
    ML-KEM~\cite{nist2024_fips203} & $\mathbb{Z}_q$ & $\mathbb{Z}_q[X]/(X^N+1)^k$ & Module-LWE \\
    ML-DSA~\cite{nist2024_fips204} & $\mathbb{Z}_q$ & $\mathbb{Z}_q[X]/(X^N+1)^k$ & Module-LWE \\
    BFV~\cite{fan2012somewhat} & $\mathbb{Z}_t$ & $\mathbb{Z}_q[X]/(X^N + 1)$ & Ring-LWE \\
    BGV~\cite{brakerski2014leveled} & $\mathbb{Z}_t$ & $\mathbb{Z}_q[X]/(X^N + 1)$ & Ring-LWE \\
    FHEW~\cite{ducas2015fhew} & $\mathbb{F}_2$ & Torus $\mathbb{T}$ & LWE over Torus \\
    CKKS~\cite{cheon2017homomorphic} & $\mathbb{C}$ (approx.) & $\mathbb{Z}_q[X]/(X^N + 1)$ & Ring-LWE \\
    TFHE~\cite{chillotti2020tfhe} & $\mathbb{F}_2$ & Torus $\mathbb{T}$ & LWE over Torus\\
    \hline
    \end{tblr}}

\end{table}

\subsection{\glsentryshort{fhew} and \glsentryshort{tfhe} -- Binary Arithmetic}
\label{subsec:fhew-tfhe}

\gls{fhew} and \gls{tfhe} operate over a binary plaintext space ($\mathbb{F}_2$) and a Torus ciphertext space, which is fundamentally incompatible with \gls{mlkem}'s integer arithmetic over $\mathbb{Z}_q$.
Furthermore, supporting both binary and prime-field characteristics in hardware requires separate datapaths.
As shown by Ras et al.~\cite{cryptoeprint:2025/601}, a unified butterfly architecture for \gls{ntt} and \gls{fafft} incurs a significant increase in silicon area. 
For these reasons, \gls{fhew} and \gls{tfhe} are excluded from consideration.
\vspace{-0.25cm}

\subsection{\glsentryshort{ckks} -- Approximate Arithmetic}
\label{subsec:ckks}

\gls{ckks} operates over complex numbers $\mathbb{C}$ with approximate decryption, making it well-suited for real-valued machine learning workloads but incompatible with the exact integer semantics required by \gls{mlkem}. 
The encoding and decoding overhead between $\mathbb{C}$ and $R_q = \mathbb{Z}_q[X]/(X^N + 1)$ introduces additional complexity without benefit in our context of average heart rate computation (see~\autoref{sec:usecase}).
\vspace{-0.25cm}

\subsection{\glsentryshort{bfv} vs. \glsentryshort{bgv}}
\label{subsec:bfv-bgv}

Both \gls{bfv} and \gls{bgv} share the same plaintext space $\mathbb{Z}_t$ and ciphertext space $R_q = \mathbb{Z}_q[X]/(X^N + 1)$, and their \gls{rlwe} security assumption directly mirrors that of \gls{mlkem}. 
The conversion from plaintext $R_t$ to ciphertext $R_q$ reduces to a simple 
modulus switch.
Kim et al.~\cite{cryptoeprint:2021/204} show that the two schemes are largely equivalent, but \gls{bgv} outperforms \gls{bfv} for small $t$ moduli and offers finer granularity through its leveled variant. We therefore select \textbf{\gls{bgv}} for the remainder of this work.
\vspace{-0.15cm}
\section{Description of \gls{bgv} Optimization}
\label{sec:bgv-optimization}

\subsection{Leveled \glsxtrlong{bgv}}
\label{subsec:bgv-def}

In this subsection, we will recall the definition of the leveled version of \gls{bgv}~\cite{brakerski2014leveled}.
We denote by $t$ and $q$ respectively the plaintext and the ciphertext modulus.
Moreover, as described by Mono et al.~\cite{cryptoeprint:2022/706}, we set $t \equiv 1 \pmod{N}$ and $q$ a chain of prime numbers, such that $q = q_{L-1} = \prod_{j=0}^{L-1} q_j$, where $q_i$ are roughly of the same size and $p_j \equiv 1 \pmod{N}$ . 
The number of primes, $L$, in the circuit is determined by its multiplicative depth, $M$.
Therefore, at any level $\ell$, we can express $q_\ell = \prod_{j=0}^{\ell} q_j$.
\vspace{-0.25cm}

\paragraph{\textbf{Key Generation.}}
\label{para:bgv-keygen}
We sample a ternary secret $s \leftarrow \chi_s$, a polynomial $a \leftarrow \mathcal{U}_{q_L}$ uniformly at random, and a Gaussian error $e \leftarrow \chi_e$.  
Define the public key $pk$ and the secret key $sk$ as $pk = (b,a) = (-a \cdot s + t\cdot e, a)$ and $sk = s \pmod{q_L}.$
\vspace{-0.25cm}

\paragraph{\textbf{Encryption.}}
\label{para:bgv-encryption}
To encrypt a message $m \in \mathcal{R}_t$ under the public key $pk = (b,a)$, we generate a ternary sample $u \leftarrow \chi_s$ and two Gaussian errors $e_0,e_1 \leftarrow \chi_e$.
We obtain a ciphertext $c = (c_0, c_1) = (b \cdot u + t \cdot e_0 + m, a \cdot u + t \cdot e_1) \pmod{q_L}$.
\vspace{-0.25cm}
\paragraph{\textbf{Decryption.}}
\label{para:bgv-decryption}
Given a ciphertext $c=(c_0,c_1)$ and a secret key $sk = s$, we compute the decrypted message as $m' = (c_0 + c_1 \cdot s \pmod{q_{\ell}}) \pmod{t}$, which recovers the message $m$ provided the error is small enough.
\vspace{-0.25cm}

\paragraph{\textbf{Homomorphic operations.}}
\label{para:operation}
Let $\alpha \in R_t$ be a constant polynomial and $c = (c_0, c_1)$, 
$c' = (c'_0, c'_1)$ two ciphertexts at the same level $\ell$. We define:
\begin{align}
\mathbf{Add}(c,c') &= (c_0 + c'_0,\ c_1 + c'_1) \pmod{q_\ell} \\
\mathbf{MulConst}(\alpha,c) &= (\alpha c_0,\ \alpha c_1) \pmod{q_\ell} \\
\mathbf{Mul}(c,c') &= (c_0 c'_0,\ c_0 c'_1 + c_1 c'_0,\ c_1 c'_1) \pmod{q_\ell}
\end{align}

where $\mathbf{Mul}$ outputs a degree-3 tuple requiring a 
\textit{relinearization} (key-switching) step to return to a standard 
two-component ciphertext.
\vspace{-0.25cm}
\paragraph{\textbf{Modulus Switching.}}
\label{para:bgv-modswitch}
Let $c = (c_0,c_1)$ be a ciphertext at level $\ell$, defined modulo $q_\ell$.
Let $q_{\ell-1} < q_\ell$ be the next modulus in the modulus chain, we define the modulus switching operation as follows: $$\mathbf{ModSwitch}(c) = \Big(\left\lfloor \tfrac{q_{\ell-1}}{q_\ell} \cdot c_0 \right\rceil, \left\lfloor \tfrac{q_{\ell-1}}{q_\ell} \cdot c_1 \right\rceil
\Big)$$

This operation preserves the encrypted message while reducing the modulus from $q_\ell$ to $q_{\ell-1}$. Moreover, it proportionally reduces the noise, up to a small additional rounding error introduced during the scaling step.\vspace{-0.25cm}
\paragraph{\textbf{Critical Quantity.}}
\label{para:critical-quantity}
Since noise accumulates with each homomorphic operation, it is important to track the noise level: beyond a certain limit, decryption fails because of excessive noise. To this end, Mono et al.~define in~\cite{cryptoeprint:2022/706} the notion of \emph{critical quantity} (or noise budget).
Let $\mathbf{c} = (c, \ell, \nu)$ be an extended ciphertext in the BGV scheme, where $c = (c_0, c_1)$ is a ciphertext pair, $\ell$ denotes the current level, and $q_\ell$ is the associated ciphertext modulus.
The \emph{critical quantity} of $c$ is the polynomial $\nu = \bigl[c_0 + c_1 \cdot s\bigr]_{q_\ell},$ obtained by evaluating the ciphertext as a polynomial in the secret key $s$.
Decryption is correct if and only if $\|\nu\|_{\infty} \leq c_m \|\nu\|_{\mathrm{can}} < \frac{q_\ell}{2},$ where $c_m$ is the ring expansion factor and $\|\cdot\|_{\mathrm{can}}$ denotes the canonical embedding norm. 
The critical quantity grows with each homomorphic operation; its bounds for the main operations are given below.
\vspace{-0.25cm}
\paragraph{\textbf{Circuit.}}
\label{para:circuit}
In a leveled \gls{fhe} scheme, an \textit{arithmetic circuit} is a graph whose nodes represent homomorphic operations (additions, multiplications, Boolean operations) applied to ciphertexts; the determining parameter is its multiplicative depth $L$, defined as the maximum number of multiplications on any path from input to output.
\vspace{-0.35cm}
\subsection{Crypto-Agility: \gls{bgv} and \gls{mlkem}}
\label{subsec:crypto-agility}

In this subsection, we explain how key generation, encryption and decryption can be shared between \gls{bgv} and \gls{mlkem}.
It is important to note that our goal is not to optimize the homomorphic operations defined in~\autoref{subsec:bgv-def}, as they are executed in the server component of the \gls{he} protocol.
Nevertheless, they still have an influence on the selection of parameters described in~\autoref{subsec:bgv-parameters}.
As discussed in~\autoref{sec:schema-selection}, both schemes operate on the same type of arithmetic.

\begin{figure}

    \centering
    \includegraphics[width=1\textwidth]{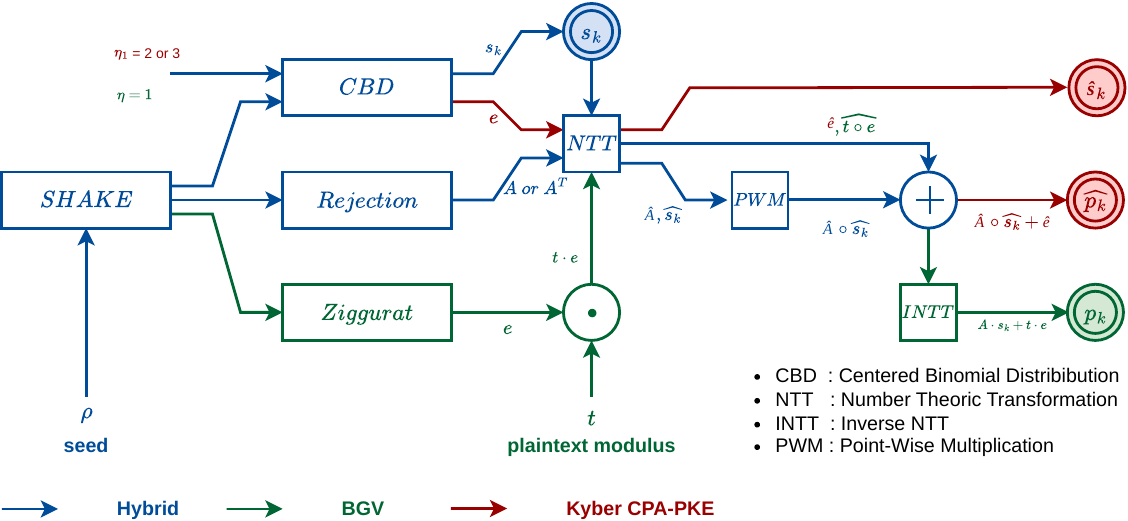}
    \caption{Hybrid key generation for \gls{bgv}/\gls{mlkem}.}
    \label{fig:BGV-Kyber-Keygen}
    \vspace{-0.65cm}
\end{figure}

\begin{figure}
    \centering
    \includegraphics[width=1\textwidth]{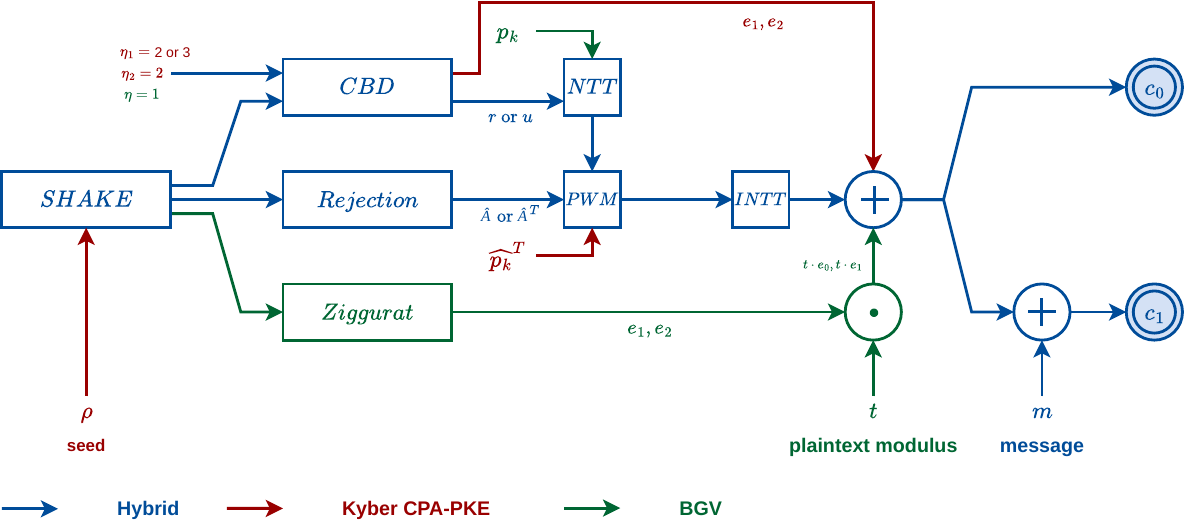}
    \caption{Hybrid encryption for \gls{bgv}/\gls{mlkem}.}
    \label{fig:BGV-Kyber-Encryption}
\end{figure}

\begin{figure}
    \centering
    \vspace{-0.75cm}
    \includegraphics[width=.7\textwidth]{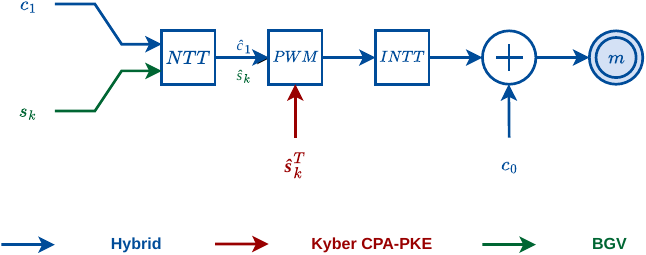}
    \caption{Hybrid decryption for \gls{bgv}/\gls{mlkem}.}
    \label{fig:BGV-Kyber-Decryption}
\end{figure}

\begin{remark}
    \label{remark:cpapke}
    Here, we use the CPA-PKE version of \gls{mlkem}~\cite{nist2024_fips203}.
    For the CPA-KEM version, we can reuse the optimizations from the CPA-PKE version, since the keys are encapsulated.
    For the sake of simplicity, \autoref{fig:BGV-Kyber-Encryption} is drawn for the case of \gls{mlkem}.
    We note that the encrypted pair $(c_0 = A^T \cdot r + e_1, c_1 = t^T \cdot r + e_2 + m)$ where $m = \mathtt{Decompress}_q(M, 1)$.
    In addition, we have omitted the $\mathtt{Compress}_q$ and $\mathtt{Decompress}_q$ stages from the diagram, since they are not part of \gls{bgv}.
    These operations can be managed independently and concurrently.
    The same approach is applied to decryption in~\autoref{fig:BGV-Kyber-Decryption}.
\end{remark}

We observe that \gls{bgv} and \gls{mlkem} share many similarities for key generation (see~\autoref{fig:BGV-Kyber-Keygen}) and encryption (see~\autoref{fig:BGV-Kyber-Encryption}).
Both algorithms involve polynomial multiplication using the \gls{ntt} and random number generation.
However, the main distinction is that, in \gls{mlkem}, the keys remain in the \gls{ntt} domain.
\autoref{fig:BGV-Kyber-Decryption} shows that both cases share identical functions.
However, like in encryption, the key must be passed in order to perform calculations in the \gls{ntt} domain.

\begin{remark}
Although the \gls{mlkem} standard keeps the keys in the \gls{ntt} domain, they could be transferred to the classical domain, as in \gls{bgv}.
This avoids having to repeat the transfer operation during encryption.
\end{remark}

\subsection{Random Number Generator}
\label{subsec:random-generator}

\paragraph{\textbf{Uniform Rejection.}}
\label{para:uniform-rejection}

In \gls{bgv}, we need to sample an element $a\in \left[-\frac{q}{2} \; ; \; \frac{q}{2}\right]$ from the uniform distribution.
The idea is to reuse the \gls{prf} or \gls{xof} defined in~\cite{nist2024_fips203} for \gls{mlkem}, which uses $\mathsf{SHAKE}$.
In general, the same function can be used for both algorithms, namely $\mathsf{SHAKE}$ (SHAKE128 as \gls{xof} and SHAKE256 as \gls{prf} in \gls{mlkem}), standardized in~\cite{nist2015_fips202}.

\paragraph{\textbf{Gaussian Sample.}}
\label{para:gaussian}
When creating an error sample ($e, e_0, e_1 \leftarrow \chi_e$) for \gls{bgv}, as suggested by the \gls{hess} in~\cite{HomomorphicEncryptionSecurityStandard}, one should use a normal distribution with a standard deviation of $\sigma 
\approx 3.19$.
There are several techniques for generating random normal numbers, including the Box-Muller, Knuth-Yao, Kinderman, Polar, Monty Python, Von Neumann-type, and Ziggurat methods. 
According to Matsumoto~\cite{matsumoto2025secure} and Prest~\cite{prest:tel-01245066}, the Ziggurat method~\cite{ziggurat} is the fastest and has the lowest rejection rate of all the previously mentioned methods.
Furthermore, since the Ziggurat method is a rejection method that requires a source of uniformly distributed random numbers, we can also use the SHAKE algorithm of \gls{mlkem} to generate them.
\vspace{-0.25cm}
\paragraph{\textbf{\gls{cbd}.}}
\label{para:cbd}
For the generation of the secret sample ($s \leftarrow \chi_s$) in \gls{bgv}, the \gls{hess}~\cite{HomomorphicEncryptionSecurityStandard} recommends a ternary distribution, i.e., selecting each coefficient uniformly in $\{-1, 0, 1\}$.
Generating a ternary distribution is equivalent to using a \gls{cbd} with $\eta = 1$, defined by:
    $$CBD_\eta:= \left (\sum_{i=1}^{\eta} a[i] -  \sum_{i=1}^{\eta} b [i]\ : \ a, b \xleftarrow {\$} \{0;1\}^{\eta} \right).$$

As can be seen in the above definition, the generation of random numbers in \gls{cbd} does not depend on the modulus $q$ or the degree $N$.
For this reason, a single component is used for both ML-KEM and BGV, with only the output length and $\eta$ changing.
SHAKE can also be used to generate the seed for the \gls{cbd} process.

\subsection{Selection of \gls{bgv} Parameters}
\label{subsec:bgv-parameters}

To minimize the computational cost, we use the leveled version of \gls{bgv}.
Considering our use case (\autoref{sec:usecase}), we observed that only one or two multiplications are executed in the homomorphic domain.
We therefore restrict ourselves to 2-level \gls{bgv} ($q = q_0 \cdot q_1$, \textit{e.g.} mean computation) and 3-level \gls{bgv} ($q = q_0 \cdot q_1 \cdot q_2$, \textit{e.g.} standard deviation), whose respective multiplicative depths exactly cover the circuits involved.
In both cases, the noise generated by additions is negligible compared to that of scalar multiplication.

To ensure a minimum level of security and an optimal noise budget for decryption, we rely on the \texttt{FHEgen} tool, as recommended in~\cite{cryptoeprint:2023/600,cryptoeprint:2022/706}.
In the first scenario, we configure \texttt{FHEgen}\footnote{Note that \texttt{FHEgen} does not offer an option for zero multiplications, and that the noise generated by additions is negligible compared to that of scalar multiplication.} with a 128-bit security level for one and two multiplication depths. In our case, we selected the moduli $q_i$ as follows:
\begin{itemize}
    \item $q_0 = 13 \cdot 2^{8} + 1 = 3\ 329$ (as \gls{mlkem}),
    \item $ q_1 = 1023 \cdot 2^{13} + 1 = 8\ 380\ 417$ (as \gls{mldsa}),
    \item $ q_2$ a prime number such that $\log_2(q_2) \approx 63$, e.g. $q_2 = 4611686160161308673$.
\end{itemize}

In line with Jyrwa et al.’s suggestion in~\cite{11324134}, we approximate $\log_2(q_0) + \log_2(q_1) \approx 35$. Therefore, $\log_2(q_2)$ must be approximately $63$ to obtain $\log_2(Q)\approx 97$ (as proposed by \texttt{FHEgen}) while maintaining the same multiplication depth.

However, \texttt{FHEgen} uses a heuristic to determine the level of security against known attacks (uSVP, BDD, and hybrid).
Therefore, we use the \texttt{fastParameterSelection} tool~\cite{biasioli_tool_2024}, which provides exact verification for each attack.
Note that it does not ensure correct decryption, which is why combining the two tools is beneficial.
In the appendix, \autoref{tab:secu-analysis} summarizes the security levels for each modulus and each attack given by \texttt{fastParameterSelection}; we can see that, in the worst-case scenario, the security level is 285, which leaves us a sufficient margin for a target security level of 128 bits.

\subsection{\gls{ntt} Computation}
\label{subsec:ntt-computation}

In this subsection, we introduce the \gls{ntt}, a finite-field version of the \gls{dft}~\cite{Agarwal1975NumberTT,Pollard1971TheFF}.
The \gls{ntt} significantly accelerates polynomial multiplication by reducing the computational complexity from quadratic to quasi-linear time.
We therefore build on existing work on the \gls{ntt} in the \gls{lwe} context to adapt it to our \gls{bgv} and \gls{mlkem} use case.

\begin{definition}[\glsxtrlong{ntt}]
Let $a\in\mathbb{Z}_q$ and $\zeta_n$ be a primitive $n$-th root of unity in $\mathbb{Z}_q$.
The forward transform, called the \textbf{\gls{ntt}}, is defined by $\hat{a} = \mathrm{NTT}(a)$, where $$\hat{a}_j = \sum_{i=0}^{n-1} a_i \, \zeta_n^{i \cdot j} \pmod q, \quad \text{for } j = 0, 1, \dots, n - 1.$$

The inverse transform, called the \textbf{\gls{intt}}, is defined by $a = \mathrm{INTT}(\hat{a})$, where $$a_i = n^{-1} \sum_{j=0}^{n-1} \hat{a}_j \, \zeta_n^{-i\cdot j} \pmod q,
\quad \text{for } i = 0, 1, \dots, n - 1.$$

The NTT and INTT satisfy $a = \mathrm{INTT}(\mathrm{NTT}(a))$, $\forall a\in\mathbb{Z}_q$.

\end{definition}

\begin{proposition}[Convolution]
Let $c = a \cdot b \in \mathbb{Z}_q[X]/(X^N - 1)$ be the cyclic convolution of $a$ and $b$. 
Then it holds that $\mathrm{NTT}(c) = \mathrm{NTT}(a) \circ \mathrm{NTT}(b),$ where $\circ$ denotes component-wise multiplication.
\end{proposition}

\paragraph{\textbf{\glsxtrlong{rns}.}}
\label{para:rns}

As mentioned in~\autoref{subsec:bgv-parameters}, at most two multiplication levels are required.
To minimize noise from homomorphic multiplication while maintaining 128-bit security in \gls{bgv}, we use a large $q$ value, unlike in \gls{mlkem}. 
In leveled \gls{bgv}, $q$ is defined as a product of primes, enabling efficient modulus switching after multiplication to reduce noise. 
This approach optimizes \gls{ntt} computation.

\begin{definition}[\gls{crt}]
\label{def:crt}
    The \gls{crt} states that if the $m_i$ are pairwise co-prime and 
    $x \equiv a_i \bmod{m_i}$ for all $i = 1, \ldots, k$, then $x$ has 
    a unique solution modulo $m_1 m_2 \cdots m_k$.
\end{definition}

Indeed, according to~\autoref{def:crt}, we can split the computation of the \gls{ntt} modulo $q$ into one \gls{ntt} per $q_i$.
We have: $$\mathbb{Z}_q[X]/(X^n + 1) \cong \mathbb{Z}_{q_{0}}[X]/(X^n + 1) \times \mathbb{Z}_{q_{1}}[X]/(X^n + 1) \times \mathbb{Z}_{q_{2}}[X]/(X^n + 1)$$
After applying \gls{ntt} to each $q_i$, we recombine to obtain the result modulo $q$.

\paragraph{\textbf{\gls{ntt} radix-2.}}
\label{para:2-radix}

In order to calculate the \gls{ntt}, the most common approach is based on \textit{divide and conquer}: the problem is recursively broken down into smaller sub-problems, a strategy known as \textit{radix-$r$}. 
The main advantages are the reduction in the degree of the polynomials manipulated at each stage and the possibility of parallelizing the calculations, which is particularly beneficial in a multi-core context.
In the case of \gls{mlkem} -- as well as \gls{mldsa} -- the reference implementation~\cite{Kyber:2023} adopts a radix-2 \gls{ntt}.

\begin{definition}[\gls{ntt} radix-2]
    An \gls{ntt} \textit{radix-$2$} is a recursive decomposition of the problem of size $n$ into two subproblems of size $n/2$, exploiting the structure of the roots of unity in $\mathbb{Z}_q$. 
For a polynomial $a \in R_q$ of degree $n-1$, we separate the even and odd coefficients:
$$
    \hat{a}[k] = \sum_{j=0}^{n-1} a[j]\, \zeta_n^{jk} = \underbrace{\sum_{j=0}^{n/2-1} a[2j]\, \zeta_{n/2}^{jk}}_{A_{\mathrm{even}}[k]} + \ \zeta_n^k \cdot \underbrace{\sum_{j=0}^{n/2-1} a[2j+1]\, \zeta_{n/2}^{jk}}_{A_{\mathrm{odd}}[k]},
$$

which reduces the calculation of an \gls{ntt} of size $n$ to two \glspl{ntt} of size $n/2$, reducing the complexity from $\mathcal{O}(n^2)$ to $\mathcal{O}(n \log n)$.
The two decompositions are the \gls{ct} butterfly~\cite{d3ea2d52-5ab2-3128-8b80-efb85267295d} and the \gls{gs} butterfly~\cite{Gentleman1966FastFT} in \autoref{fig:ct-gs}, defined respectively by:

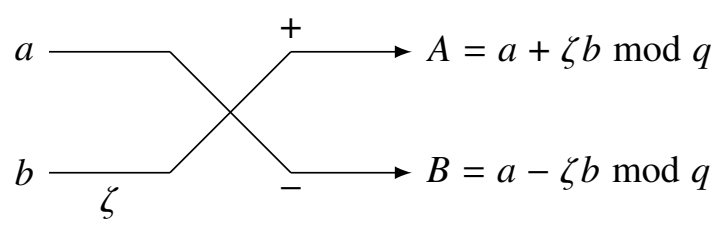
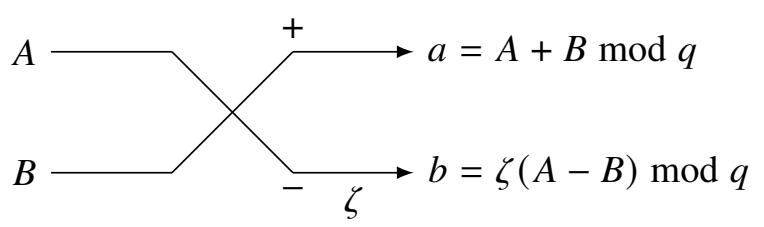
\begin{figure}
\centering
\scalebox{0.8}{%
\begin{subfigure}[b]{0.5\textwidth}
\centering
\begin{tikzpicture}[>=latex, font=\small]
  \node[left] at (0,1)  {$a$};
  \node[left] at (0,0)  {$b$};
  \draw (0,1) -- (1,1);
  \draw (0,0) -- (1,0) node[midway, below] {$\zeta$};
  \draw (1,1) -- (2,0);
  \draw (1,0) -- (2,1);
  \node[above] at (2,1) {$+$};
  \node[below] at (2,0) {$-$};
  \draw[->] (2,1) -- (3,1);
  \draw[->] (2,0) -- (3,0);
  \node[right] at (3,1) {$A = a + \zeta b \bmod q$};
  \node[right] at (3,0) {$B = a - \zeta b \bmod q$};
\end{tikzpicture}
\caption{Butterfly CT}
\end{subfigure}
\hfill
\begin{subfigure}[b]{0.5\textwidth}
\centering
\begin{tikzpicture}[>=latex, font=\small]
  \node[left] at (0,1)  {$A$};
  \node[left] at (0,0)  {$B$};
  \draw (0,1) -- (1,1);
  \draw (0,0) -- (1,0);
  \draw (1,1) -- (2,0);
  \draw (1,0) -- (2,1);
  \node[above] at (2,1) {$+$};
  \node[below] at (2,0) {$-$};
  \draw[->] (2,1) -- (3,1);
  \draw[->] (2,0) -- (3,0) node[midway, below] {$\zeta$};
  \node[right] at (3,1) {$a = A + B \bmod q$};
  \node[right] at (3,0) {$b = \zeta(A - B) \bmod q$};
\end{tikzpicture}
\caption{Butterfly GS}
\end{subfigure}
}
\caption{Butterfly CT/GS}
\label{fig:ct-gs}
\vspace{-0.5cm}
\end{figure}
where $\zeta$ is a root of unity in $\mathbb{Z}_q$. 
The \gls{ct} butterfly is mainly used in the forward direction (\gls{ntt})~\cite{d3ea2d52-5ab2-3128-8b80-efb85267295d}, while the \gls{gs} butterfly is mainly used in the inverse direction (\gls{intt}); their composition ensures that $\gls{intt} \circ \gls{ntt} = \mathrm{Id}$.
\end{definition}

We recognize that the prime $q$ of \gls{mlkem} is constructed in the form $q = k \cdot 2^m + 1$, so that $\mathbb{Z}_q$ contains primitive $2^m$-th roots of unity. This allows for faster computations using NTTs. 
For leveled \gls{bgv}, $Q = \prod_{i=0}^{L-1} q_{i}$, where each $q_{i}$ is a small prime.
In order to optimize with \gls{mlkem}, the idea is to take $q_{i}$ to be of the form $q_{i} = k\cdot 2^m + 1$.
As a result, the same accelerator used by \gls{mlkem} can be used for NTTs.

We want to find the \gls{ntt} representation $NTT(a) \in \mathbb{Z}_{q_{i}}[X]/(X^{8192} + 1)$ for each $q_i$.
Moreover, for $q_0 = 3\ 329$, $N = 8\ 192$ does not divide $q_0 - 1$, so there is no primitive $N$-th root of unity in $\mathbb{Z}_{q_0}$.

We propose a fast decomposition technique that factors the polynomial ring into a $32 \times 256$ product, allowing reuse of an \gls{ntt} module of size $n = 256$. 
However, this approach makes it impossible to directly apply 2-radix \gls{ntt} to compute the \gls{ntt} on the quotient ring: $$\mathbb{Z}_{q_i}[X]/(X^n + 1) \cong \mathbb{Z}_{q_i}[X]/(X^{n_1} + 1) \times \mathbb{Z}_{q_i}[X]/(X^{n_2} + 1),$$ where $n = n_1 \cdot n_2$ with $n_1 = 32$ and $n_2 = 256$.
To address this issue, we consider several options:
\begin{enumerate}
    \item \textbf{Increase $q_i$:} A larger parameter $q$ can provide a primitive root of unity. However, this approach carries a substantial memory cost, as it requires storing or recalculating the unique roots of unity associated with the new $q_i$.
    \item \textbf{Decrease $n$:} Decreasing the parameter $n$ in the BGV scheme would lower its security, which is not an acceptable option.
    \item \textbf{Increase $n$:} Increasing the parameter $n$ leads to a higher number of coefficients to process, increasing both storage requirements and algorithmic complexity.
    \item \textbf{Use another algorithm:} Compute the \gls{ntt} with a different approach.
\end{enumerate}
Therefore, we decided to use \emph{Nussbaumer's trick} to split the \gls{ntt}.

\paragraph{\textbf{Nussbaumer's Trick \gls{ntt}.}}
\label{para:nussbaumer}

Unlike the Four-Step NTT, Nussbaumer's trick does not require a primitive root of unity of order $N = 8\ 192$. 
Instead, only roots of unity for the smaller transforms of size $2n = 256$ are needed. 

\begin{definition}[Nussbaumer's trick~\cite{bernstein2001multidigit,1163372}]
\label{def:nussbaumertrick}
    Let $N = 2nm$ be the polynomial degree. 
    Nussbaumer's trick defines the ring isomorphism via the mapping
\begin{equation}
    \label{eq:nussbaumer}
    \begin{aligned}
        \Psi : \mathbb{Z}_q[x]/(x^{2nm} + 1) &\xrightarrow{\cong} \left(\mathbb{Z}_q[y]/(y^{2n} + 1)\right)[x]/(x^m - y) \\[6pt]
        \sum_{i=0}^{N - 1} a_i \cdot x^i 
        &\mapsto 
        \sum_{i=0}^{m-1} \underbrace{\left(\sum_{j=0}^{2n-1} a_{m\cdot j + i} \cdot y^j \right)}_{A_j \in \mathbb{Z}_q[y]/(y^{2n} + 1)} x^i, \text{ with } y = x^m.
    \end{aligned}
\end{equation}
\end{definition}

\begin{remark}
Note that $y$ should be the primitive $4n$-th root of unity in $\mathbb{Z}_q[y]/(y^{2n} + 1)$ to be \gls{ntt}-friendly.
\end{remark}

To improve efficiency, we can rewrite~\autoref{eq:nussbaumer} as a sum of $m$ instances of the \gls{ntt} on $\mathbb{Z}_q[y]/(y^{2n} + 1)$, which, in our case, yields the following formulation:

Let $N = 8\ 192 = 32 \times 256$. 

\begin{enumerate}
    \setlength\itemsep{1.5ex}
    \item \textbf{Partitioning the coefficient vector.}  
    We can split the vector $(a_i)$ into $32$ sub-vectors of $256$ elements such that: 
    \begin{equation}
        \label{eq:partitioning-ntt}
        a^{(i)} = \big(a_i,\, a_{i + 32},\, a_{i + 2 \cdot 32},\, \dots,\, a_{i+ 255 \cdot 32}\big), \quad \text{for }i = 0, \dots, 31.
    \end{equation}

    \item \textbf{Applying the $NTT_{256}$ to each sub-vector.}  
    For each $i$, we can compute $$\widehat{a^{(i)}} = \mathrm{NTT}_{256}\big(a^{(i)}\big) = \Big( \sum_{j=0}^{2n-1} a_{i+m j} \, \zeta^{jk} \Big), 
    \quad \text{for }k = 0, \dots, 255.$$
    where $\zeta$ is a primitive $256$-th root of unity modulo $q$.
    \item \textbf{Final expression in NTT form.}  
    Replacing each inner sum with its NTT version, the global transform becomes
    \begin{equation}
        \label{eq:final-ntt}
        \sum_{i=0}^{N-1} a_i \, x^i \ \ \longmapsto\;\;
    \sum_{i=0}^{m-1} \Big( \sum_{k=0}^{2n-1} \widehat{a}_k^{(i)} \, y^k \Big) x^i,
    \end{equation}
\end{enumerate}

This trick allows us to use the same \gls{ntt} function for both cases (\gls{mlkem} and \gls{bgv}); it can be the reference implementation from the \gls{nist} standard~\cite{nist2024_fips203} or any other optimized implementation.
In addition, the fact that we have selected $q_0 = 3\  329$ for \gls{bgv} allows us to reuse the $\zeta$ table from \gls{mlkem}\footnote{The same reasoning applies to \gls{mldsa}.}.
In \autoref{tab:complexity}, we summarize the complexity of various multiplication algorithms.
\section{Implementation}
\label{sec:implementation}

In order to show the feasibility of our proposed scheme and to test the selected parameters with our use case, we have implemented the embedded part (key generation, encryption and decryption) on a PolarFire SoC Discovery Kit platform and the server part (addition and multiplication) on an Intel Xeon Gold 6226R with SageMath.

\subsection{Parameter Selection}
\label{subsec:param-implem}

According to \cite{Shaffer}, the literature distinguishes several time windows for the analysis of interbeat intervals: an ultra-short window of 60 seconds ($\approx$60–80 $toc$), a window of 240 seconds ($\approx$240–320 $toc$), and the standard 5-minute window ($\approx$300-400 $toc$).
 For long-term analyses, recordings can extend up to 24 hours ($\approx$100,000 $toc$), particularly for calculating the \gls{sdnn}.
To ensure that the chosen parameters allow accurate decryption, we evaluate the critical noise level in a \texttt{SageMath}\footnote{\texttt{SageMath} is a free mathematical software system based on Python~\cite{sagemath}.} implementation using the parameters from \autoref{subsec:bgv-parameters}. We do this for a standard 5-minute window of $N = 400$ ciphertexts. After $N = 400$ homomorphic additions, we measure $\|\nu\|_\infty \approx 2^{66} \;<\; B_{\max} \approx \frac{q}{2t} = 2^{79}$, which satisfies the correctness condition, leaving a margin of $13$ bits before decryption failure.
This confirms that the scheme remains correct well beyond the target
workload, supporting up to ${\approx}23\,000$ additions under these parameters.
Further details on the runtime performance of each operation are reported
in \autoref{tab:benchmark_server}.

\begin{table}[!ht]
\vspace{-0.75cm}
    \centering
        \caption{Benchmark of the mean computation with BGV on an Intel Xeon Gold 6226R
             (2.90~GHz), $n = 8192$,
             $q \approx 2^{96}$, $t = 65537$, $\sigma = 3.2$.
             Timings reported as min. over 10 runs.}
    \label{tab:benchmark_server}
    \resizebox{1\textwidth}{!}{\begin{tblr}{
        width=\textwidth,
        colspec = {lrrrr},
        row{1} = {font=\bfseries},
        row{8} = {font=\itshape},
        row{9} = {font=\itshape},
        rowsep=1pt
    }
        \hline
         & \SetCell[c=2]{c} $N = 400$ & & \SetCell[c=2]{c} $N = 10000$  & \\
        Operation & Min (ms) & Min cycles & Min (ms) & Min cycles \\
        \hline
        KeyGen    & 152.33 & $548.4 \times 10^{6}$ & 138.07 & $538.5 \times 10^{6}$ \\
        KeySwitch & 147.76 & $531.9 \times 10^{6}$ & 139.69 & $544.8 \times 10^{6}$ \\
        Encrypt   & 125.46 & $451.7 \times 10^{6}$ & 119.21 & $464.9 \times 10^{6}$ \\
        Add       &  35.87 & $129.1 \times 10^{6}$ &  38.11 & $148.6 \times 10^{6}$ \\
        Decrypt   &  27.98 & $100.7 \times 10^{6}$ &  28.58 & $111.5 \times 10^{6}$ \\
        \hline
        Encrypt (total) & \SetCell[c=2]{c} $61.88$ s \; ($154.71$ ms/op) &
                        & \SetCell[c=2]{c} $1454.25$ s \; ($145.43$ ms/op) & \\
        Add (total)     & \SetCell[c=2]{c} $19.56$ s \; ($49.14$ ms/op) &
                        & \SetCell[c=2]{c} $513.65$ s \; ($51.38$ ms/op) & \\
        \hline
    \end{tblr}}
    \vspace{-0.70cm}
\end{table}

\subsection{\gls{ntt} Implementation}
\label{subsec:ntt-implem}

In this subsection, we present the implementation of $NTT_{8192}$.
As explained previously (cf.~\autoref{sec:bgv-optimization}), one of the main ideas is to reuse as much as possible of the reference code for $NTT_{256}$ from \gls{mlkem} available in the official repository of the \textit{Crystals-Kyber} project~\cite{Kyber:2023}.
To multiply two polynomials $f, g \in \mathbb{Z}_q[x]/(x^{8192} + 1)$ using the optimization described in~\autoref{subsec:ntt-computation}, we proceed as follows:

\begin{enumerate}
    \setlength\itemsep{1.5ex}
    \item \textbf{Incomplete \gls{ntt} Transformation:} We compute the incomplete \gls{ntt} of $f$ and $g$ using the transformation $\psi_{8192}$ defined in \autoref{alg:psi}: $F = \texttt{psi\_8192}(f, \boldsymbol{\zeta}, q)$ and $G = \texttt{psi\_8192}(g, \boldsymbol{\zeta}, q)$.
    This function rearranges the coefficients of $f$ and $g$ according to the partitioning scheme in~\autoref{eq:partitioning-ntt}, then applies the \texttt{ntt\_256} function to each of the 256-coefficient sub-blocks, as described in~\cite{Kyber:2023}.
    The list of required roots of unity $\boldsymbol{\zeta}$ can be precomputed and stored in memory or hard-coded.
    
    \resizebox{0.95\textwidth}{!}{\footnotesize
    \begin{algorithm}[H]
    \caption{$\texttt{psi\_8192}(f, \boldsymbol{\zeta}, q)$ -- The forward transform $\psi_{8192}$}
    \label{alg:psi}
    \small
    \KwIn{$f \in \mathbb{Z}_q[X]/(X^{8192} + 1)$ with underlying $256$-point $\text{NTT}_{256}$}
    \KwOut{$\hat{f} := \{\hat{F}_i\}_{i=0}^{31}$}
    \BlankLine
    Map $f$ into $\Psi(f) = \sum_{j=0}^{31} F_j x^j$ where each $F_j$ is a polynomial in $\mathbb{Z}_q[y]$;
    \For{$i \leftarrow 0$ \KwTo $31$}{
        $F_i \leftarrow [0]_{j=0}^{255}$\;
        \For{$j \leftarrow 0$ \KwTo $255$}{
            $s \leftarrow 32 \cdot j + i$\;
        }
        $\hat{F}_i \leftarrow \texttt{NTT\_256}(F_i, \boldsymbol{\zeta}, q)$\;
    }
    \Return $\{\hat{F}_i\}_{i=0}^{31}$\;
    \end{algorithm}
    }
   
    \item \textbf{Negacyclic \gls{pwm}:} As specified in \autoref{alg:conv-full}, we compute $H = \texttt{conv\_8192}(F, G, \boldsymbol{\zeta}, q)$. 
    This function performs negacyclic polynomial multiplication on each 256-coefficient sub-block using the \texttt{poly\_basemul} function from~\cite{Kyber:2023}.
    After multiplication, we apply the reduction $X^{32} - Y$ when the index $s$ is at least $32$.

    \resizebox{0.95\textwidth}{!}{\footnotesize
    \begin{algorithm}[H]
    
    \caption{$\texttt{conv\_8192}(F, G, \boldsymbol{\zeta}, q)$ -- Negacyclic Convolution}
    \label{alg:conv-full}
    \KwIn{$F, G \in \mathbb{Z}_q[X]/(X^{8192}+1)$, $\boldsymbol{\zeta}$ roots, $q$ modulus}
    \KwOut{$H := F \odot G$}
    Initialize $H \gets [\,[0]^{256}\,]^{32}$\;
    \For{$j_1 \gets 0$ \KwTo $31$}{
        \For{$j_2 \gets 0$ \KwTo $31$}{
            $s \gets j_1 + j_2$\;
            $j \gets s \bmod 32$\;
            $\mathtt{tmp} \gets \texttt{poly\_basemul}(F[j_1], G[j_2], \boldsymbol{\zeta}, q)$\;
            \If{$s \geq 32$ \tcp*[r]{$\bmod{(X^{32} - Y)}$}}{
                \For{$i \gets 0$ \KwTo $63$}{
                    $z \gets \boldsymbol{\zeta}[64 + i]$\;
                    $a_0, a_1 \gets \mathtt{tmp}[4i],\ \mathtt{tmp}[4i+1]$\;
                    $a_2, a_3 \gets \mathtt{tmp}[4i+2],\ \mathtt{tmp}[4i+3]$\;
                    $\mathtt{tmp}[4i]   \gets (a_1 \cdot z) \bmod q$\;$\mathtt{tmp}[4i+1] \gets a_0$\; 
                    $\mathtt{tmp}[4i+2] \gets (-a_3 \cdot z) \bmod q$\;$\mathtt{tmp}[4i+3] \gets a_2$\;    
                }
            }
            \For{$k \gets 0$ \KwTo $255$}{
                $H[j][k] \gets (H[j][k] + \mathtt{tmp}[k]) \bmod q$\;
            }
        }
    }
    \Return $H$\;
    \end{algorithm}
    }

    \item \textbf{\gls{intt} Transformation:} We compute the inverse transformation $\psi^{-1}_{8192}$ defined in \autoref{alg:psi-inv}: $h = \texttt{psi\_inv\_8192}(H, \boldsymbol{\zeta}, q) = f \cdot g.$

    This function computes the inverse \gls{ntt} for each 256-coefficient sub-block using the \texttt{iNTT\_256} function from~\cite{Kyber:2023}, then reorders the coefficients to obtain the final multiplication result.

    \resizebox{0.95\textwidth}{!}{\footnotesize
    \begin{algorithm}[H]
    \small
    \caption{$\texttt{psi\_inv\_8192}(H, \boldsymbol{\zeta}, q)$ -- The inverse transform $\psi^{-1}_{8192}$}
    \label{alg:psi-inv}
    \small
    \KwIn{$H := \{\hat{H}_i\}_{i=0}^{31}$ with underlying 256-point $\text{iNTT}_{256}$}
    \KwOut{$h \in \mathbb{Z}_q[X]/(X^{8192} + 1)$ where $h := \{h_k\}_{k=0}^{8191}$}
    \BlankLine
    Initialize $h \leftarrow [0]_{k=0}^{8191}$\;
    \For{$i \leftarrow 0$ \KwTo $31$}{
        $H_i \leftarrow \texttt{iNTT}_{256}(\hat{H}_i, \boldsymbol{\zeta}, q)$\;
        \For{$j \leftarrow 0$ \KwTo $255$}{
            $\text{idx} \leftarrow i + 32 \cdot j$\;
            $h[\text{idx}] \leftarrow H_i[j]$\;
        }
    }
    \Return $h$\;
    \end{algorithm}
    }

\end{enumerate}

\subsection{Embedded Implementation}

For the embedded implementation, we selected the PolarFire SoC Discovery Kit board from Microchip, which contains five 64-bit RISC-V cores. 
We use only a single core of the PolarFire for the benchmarks in \autoref{tab:benchmark_ntt} and \autoref{tab:benchmark_bgv_riscv}.
This benchmark confirms Nussbaumer's structure: the \texttt{psi\_8192} and \texttt{psi\_inv\_8192} operations, which correspond to multiplications by monomials (simple rotations of coefficients), are cheap ($\approx 8\,\%$ and $\approx 6\,\%$ of the total time, respectively), while \texttt{conv\_8192} accounts for $\approx 85\,\%$ of the computation time.

Each multiplication modulo $q$ requires a reduction using native integer division, which can be expensive.
These reductions could be replaced with Montgomery reduction, which replaces costly divisions with multiplications and bit shifts, significantly reducing the dominant cost of the \texttt{conv\_8192} function.
Furthermore, we observed that \gls{bgv} key generation using Nussbaumer’s trick takes 94\,679\,558 cycles, or $157.80\,\mu$s.

Note that our implementation also works for $q  = 8380417$ (\gls{mldsa}) with performance identical to that of $q = 3329$.
Furthermore, it shares the same implementation, up to a few constants (roots of unity).

\begin{table}[ht]
\centering
\caption{Benchmark of NTT computation on a RISC-V U54 PolarFire SoC Discovery Kit with \texttt{-O3} option (600\,MHz), for $n=8192$, $q = 3329$ and $q = 8380417$.}
\label{tab:benchmark_ntt}
\begin{tblr}{
  width=\textwidth,
  colspec={X[l] X[r] X[r]},
  row{1}={font=\bfseries},
  rowsep=1pt
}
\hline
Operation & Max. clock cycles & Max. time (ms) \\
\hline
\texttt{psi\_8192}      & 4\,073\,114  & 6.79  \\
\texttt{conv\_8192}     & 41\,265\,659 & 68.78 \\
\texttt{psi\_inv\_8192} & 2\,959\,048  & 4.93  \\
\texttt{add\_8192}      & 245\,801     & 0.41  \\
\hline
\end{tblr}
\end{table}

\begin{table}[ht]
\centering
\caption{Benchmark of the BGV implementation on a RISC-V U54 PolarFire SoC Discovery Kit with \texttt{-O3} option (600\,MHz), for $n=8192$, $q = 3329$ and $q = 8380417$.}
\label{tab:benchmark_bgv_riscv}
\begin{tblr}{
  width=\textwidth,
  colspec={X[l] X[r] X[r] X[r] X[r]},
  row{1}={font=\bfseries\small},
  rowsep=1pt
}
\hline
Operation & Max. clock cycles & Max. time (ms) & Code size (KB) & Memory usage (KB) \\
\hline
BGV KeyGen   & 56\,235\,000 & 93.73  & 47.95 & 567.92 \\
BGV Encrypt  & 94\,679\,558 & 157.80 & 48.42 & 699.97 \\
BGV Decrypt  & 49\,602\,603 & 82.67  & 48.03 & 567.98 \\
\hline
\end{tblr}
\end{table}
\section{Conclusion and Future Work}
\label{sec:conclusion}

In this paper we have explored the possibilities for an agile cryptographic scheme that combines \gls{mlkem} with \gls{fhe} in an embedded setting for the secure processing of health data.
We have shown how large parts of the algorithms can be shared with the goal of minimizing the implementation's code size and memory footprint, such as the random number generators and the primitive roots. Choosing \gls{bgv} as the homomorphic framework further allows the use of common arithmetic computations. 
However, the fusion of the parameters entails some challenges, in particular for the implementation of the \gls{ntt}.
In addition, our implementation trick allows us to reuse the hardware accelerator of an existing \gls{mlkem} implementation to add a \gls{bgv} implementation.
We have demonstrated with a proof-of-concept implementation that our solution is sound and effective for typical computations in a health data case study.

As a next step in our work, we plan to explore different trade-offs between optimized software implementations (e.g., using multiple cores) and the use of hardware accelerators.
There is a large design space regarding the algorithmic choices for the involved arithmetic computations, and they need to be adapted to the target platform. Finally, in an embedded setting, side-channel attacks should be taken into account.
In future work, we would like to study the vulnerability of our co-design to such attacks and propose side-channel resistant implementations.
\appendix
\section*{Appendix}
\label{sec:appendix}


\begin{table}[H]
\centering
\caption{Security analysis using \texttt{fastParameterSelection} for an LWE dimension of 8192}
\label{tab:secu-analysis}
\resizebox{1\textwidth}{!}{
\begin{tblr}{
    width=\textwidth,
    colspec = {l *{9}{r}}, 
    row{1} = {font=\bfseries}, 
    rowsep=1pt
}
\hline
    Q & Log Q & USVP & $\text{USVP}_s$ & USVP num & BDD & $\text{BDD}_s$ & BDD num & Output \\
\hline
    $q_0$ & 12 & 2737 & 4159 & 2673 & 2788 & 3698 & 2696 & 3698 \\
    $q_0q_1$ & 35 & 948 & 1145 & 1009 & 1009 & 1063 & 1010 & 1063 \\
    $q_0q_1q_2$ & 97 & 285 & 330 & 320 & 320 & 322 & 320 & 322 \\
\hline
\end{tblr}}
\end{table}


\begin{table}[H]
\centering
\caption{Comparison of polynomial multiplication algorithms in
$\mathbb{Z}_q[X]/(X^n+1)$ for $n=8192$.}
\label{tab:complexity}
\resizebox{\textwidth}{!}{\begin{tblr}{
width=\textwidth,
colspec={lllr},
row{1}={font=\bfseries},
rowsep=1pt
}
\hline
Algorithm & Complexity & Condition & Mult. for $n=8192$ \\
\hline
Schoolbook &
$O(n^2)$ &
None &
$67\,108\,864$ \\
Karatsuba~\cite{Kartasuba} &
$O(n^{1.585})$ &
None &
$\approx 1.84 \times 10^{6}$ \\
Toom--Cook 3~\cite{98a25de6-cb12-3a2a-b845-a3c85fecf399} &
$O(n^{1.465})$ &
None &
$\approx 6.9 \times 10^{5}$  \\

Radix-2 NTT (GS) \cite{d3ea2d52-5ab2-3128-8b80-efb85267295d} &
$O(n \log n)$ &
$q \equiv 1 \pmod{2n}$ &
$53\,248$  \\
Radix-2 NTT (CT) \cite{d3ea2d52-5ab2-3128-8b80-efb85267295d}  &
$O(n \log n)$ &
$q \equiv 1 \pmod{2n}$ &
$53\,248$ \\
Nussbaumer \cite{1163372} &
$O(n \log n)$ &
None &
$32\,768$ \\
\hline
\end{tblr}}
\end{table}


\begin{figure}[H]
    \centering
    \begin{minipage}{0.48\linewidth}\begin{lstlisting}[basicstyle=\footnotesize\ttfamily]
Generated your BGV configuration !
model:   Base
sec:     189
n:       4096
t:       65537
log(q):  65 (61, 4, 4)
log(P):  10\end{lstlisting}
            \vspace{-0.1cm}
        \subcaption{One multiplication depth}
            \vspace{-0.1cm}  
        \label{fig:fhegen-term-1}
    \end{minipage}
    \hfill
    \begin{minipage}{0.48\linewidth}
        \begin{lstlisting}[basicstyle=\small\footnotesize\ttfamily]
Generated your BGV configuration !
model:   Base
sec:     210
n:       8192
t:       65537
log(q):  98 (63, 31, 4)
log(P):  38\end{lstlisting}
            \vspace{-0.1cm}
        \subcaption{Two multiplication depth}
            \vspace{-0.1cm}  
        \label{fig:fhegen-term-2}
    \end{minipage}  
    \caption{Example of \texttt{FHEgen} execution (terminal output)}
    \label{fig:fhegen-term}
    \vspace{-.5cm}
\end{figure}



\begin{algorithm}[H]
\caption{\texttt{BaseMul}$(a_0, a_1, b_0, b_1, z, q)$ \cite{Kyber:2023}}
\label{alg:basemul}
\KwIn{$a_0, a_1, b_0, b_1, z \in \mathbb{Z}_q$}
\KwOut{$(c_0, c_1)$ such that $c = a \cdot b$ in $\mathbb{Z}_q[x]/(x^2-z)$}
\BlankLine
$t \gets a_1 \cdot b_1$\;
$t \gets t \cdot z$\;
$c_0 \gets a_0 \cdot b_0 + t \bmod{q}$\;
$c_1 \gets a_0 \cdot b_1 + a_1 \cdot b_0 \bmod{q}$\;
\Return $(c_0, c_1)$\;
\end{algorithm}

\begin{algorithm}[H]
\caption{\texttt{PolyMul}$(a, b, \boldsymbol{\zeta}, q)$ -- Negacyclic \gls{pwm} \cite{Kyber:2023}}
\label{alg:poly-basemul}
\KwIn{$f, g$ two polynomials of degree 256}
\KwOut{$h = f \cdot g$}
\BlankLine
$\hat{f} \gets \texttt{NTT\_256}(f, \boldsymbol{\zeta}, q)$\;
$\hat{g} \gets \texttt{NTT\_256}(g, \boldsymbol{\zeta}, q)$\;
\BlankLine
\For{$i \gets 0$ \KwTo $127$}{
    $z \gets \boldsymbol{\zeta}[64 + i]$\;
    $(\hat{h}[4i],\ \hat{h}[4i+1]) \gets \texttt{BaseMul}(\hat{f}[4i], \hat{f}[4i+1], \hat{g}[4i], \hat{g}[4i+1], z, q)$\;
    $(\hat{h}[4i+2],\ \hat{h}[4i+3]) \gets \texttt{BaseMul}(\hat{f}[4i+2], \hat{f}[4i+3], \hat{g}[4i+2], \hat{g}[4i+3], -z, q)$\;
}
\BlankLine
$h \gets \texttt{NTT}^{-1}(\hat{h})$\;
\Return $h$\;
\end{algorithm}

\begin{algorithm}[H]
\caption{\texttt{NTT\_256}$(f, \boldsymbol{\zeta}, q)$ -- Number Theoretic Transform \cite{Kyber:2023}}
\label{alg:ntt-256}
\KwIn{$f = [f_0, \ldots, f_{255}]$ (array of 256 integers)}
\KwOut{$F = \texttt{NTT}_{256}(f)$}
\BlankLine
$k \gets 1$\;
\For{$\text{len} \gets 128$; $\text{len} \geq 2$; $\text{len} \gets \text{len} \gg 1$}{
    \For{$\text{start} \gets 0$; $\text{start} < 256$; $\text{start} \gets j + \text{len}$}{
        $z \gets \boldsymbol{\zeta}[k\mathtt{++}]$\;
        \For{$j \gets \text{start}$ \KwTo $\text{start} + \text{len} - 1$}{
            $t \gets z * F[j + \text{len}] \pmod{q}$\;
            $F[j + \text{len}] \gets F[j] - t$\;
            $F[j] \gets F[j] + t$\;
        }
    }
}
\Return $F$\;
\end{algorithm}

\begin{algorithm}[H]
\caption{\texttt{iNTT\_256}$(F, \boldsymbol{\zeta}, q)$ -- Inverse NTT \cite{Kyber:2023}}
\label{alg:intt-256}
\KwIn{$F = [F_0, \ldots, F_{255}]$}
\KwOut{$f$}
\BlankLine
$\text{const } \text{inv} \gets 128^{-1} \pmod{q}$\;
$k \gets 127$\;
\BlankLine
\For{$\text{len} = 2, 4, 8, \ldots, 128$}{
    \For{$\text{start} = 0$ \KwTo $255$ \textbf{step} $2 \cdot \text{len}$}{
        $z \gets \boldsymbol{\zeta}[k--]$\;
        \For{$j = \text{start}$ \KwTo $\text{start} + \text{len} - 1$}{
            $t \gets f[j]$\;
            $f[j] \gets (t + f[j + \text{len}]) \bmod{q}$\;
            $f[j + \text{len}] \gets f[j + \text{len}] - t$\;
            $f[j + \text{len}] \gets (z \cdot f[j + \text{len}]) \bmod{q}$\;
        }
    }
}
\BlankLine
\For{$j = 0$ \KwTo $255$}{
    $f[j] \gets (f[j] \cdot \text{inv}) \bmod q$\;
}
\BlankLine
\Return $f$\;
\end{algorithm}

%
%

\begin{credits}
\subsubsection{\ackname}
This study was funded by the “Futur \& Ruptures 2024” program managed by the IMT, Carnot TSN Institute, and the Mines-Télécom Foundation.

\subsubsection{\discintname}
The authors have no competing interests to declare that are relevant to the content of this article.
\end{credits}
%
%
%
\bibliographystyle{splncs04}
\bibliography{mybibliography}

\end{document}